\documentclass[pra,aps,showpacs]{revtex4}
\usepackage{graphics}

\newcommand*{\erfc}{\mathop{\mathrm{erfc}}}
\newcommand*{\re}{\mathop{\mathrm{Re}}}

\begin{document}

\title{Weak measurement of arrival time}

\author{J. Ruseckas}
\email{ruseckas@itpa.lt}
\author{B. Kaulakys}
\affiliation{Institute of Theoretical Physics and Astronomy,\\
 A. Go\v{s}tauto 12, 2600 Vilnius, Lithuania}
\date{\today{}}

\begin{abstract}
  The arrival time probability distribution is defined by analogy with the
  classical mechanics. The difficulty of requirement to have the values of
  non-commuting operators is circumvented using the concept of weak
  measurements. The proposed procedure is suitable to the free particles and to
  the particles subjected to an external potential, as well. It is shown that
  such an approach imposes an inherent limitation to the accuracy of the arrival
  time determination.
\end{abstract}

\maketitle

\section{Introduction}

Time plays a special role in quantum mechanics. Unlike other observables, time
remains a classical variable. It is well-known, that the self-adjoint operator
of time does not exist for bounded Hamiltonians. The problems with the time rise
from the fact that in quantum mechanics many quantities cannot have definite
values simultaneously.  However, the detection of the particles in
time-of-flight and coincidence experiments are common, and quantum mechanics
should give the method for the calculation of the arrival time. The arrival time
distribution may be useful solving the tunneling time problem, as well.
Therefore, the quantum description of the arrival time has attracted much
attention
\cite{AharBohm,Kijowski,Muga1,Baute1,Aoki,Baute2,Baute3,Baute4,Delgado,Egusquiza,Grot,Kochanski,Leon,Muga2,review}.

Aharonov and Bohm introduced the arrival time operator \cite{AharBohm}
\begin{equation}
\hat{T}_{\mathrm{AB}}=\frac{m}{2}\left((X-\hat{x})\frac{1}{\hat{p}}+
\frac{1}{\hat{p}}(X-\hat{x})\right).
\end{equation}
By imposing several conditions (normalization, positivity, minimum variance, and
a symmetry with respect to the arrival point $X$) a quantum arrival time
distribution for the free particle has been obtained by Kijowski
\cite{Kijowski}.  The Kijowski's distribution may be associated with the
positive operator valued measure generated by the eigenstates of
$\hat{T}_{\mathrm{AB}}$.  However, Kijowski's set of conditions cannot be
applied in a general case \cite{Kijowski}.
Nevertheless, the arrival time operators can be built even if the particle is
not free \cite{review,Brunetti}.

In this paper we take another approach. Since the mean arrival time even in the
classical mechanics can be infinite or the particle may not arrive at all, it is
convenient to deal not with the mean arrival time and corresponding operator
$\hat{T}$, but with the probability distribution of the arrival time. The
probability distribution of the arrival time can be obtained from the suitable
classical definition.  The non-commutativity of the operators in quantum
mechanics is circumvented by using the concept of weak measurements. Such an
approach has several advantages.  It gives, in principle, the procedure of
measuring the obtained quantity. Since in the classical mechanics all quantities
can have definite values simultaneously, weak measurements give the correct
classical limit. The concept of weak measurements has been already applied to
the time problem in quantum mechanics \cite{Ruseckas1}.

We proceed as follows. In Sec.~\ref{sec:classic} we discuss various definitions
of the arrival time in the classical mechanics. The weak measurement of the
quantum arrival time distribution is presented in Sec.~\ref{sec:weakmeas}. The
properties of the obtained quantity are analyzed in Sec.~\ref{sec:free}. Section
\ref{sec:concl} summarizes our findings.

\section{\label{sec:classic}Arrival time in classical mechanics}

In classical mechanics the particle moves along the trajectory $H(x,p)=const$ as
$t$ increases. This allows us to work out the time of arrival at the point
$x(t)=X$, by identifying the point $(x_{0},p_{0})$ of the phase space where the
particle is at $t=0$, and then following the trajectory that passes by this
point, up to the arrival at the point $X$.  If multiple crossings are possible,
one may define a distribution of arrival times with contributions from all
crossings, when no distinction is made between first, second and $n$th arrivals.
In this article we will consider such a distribution.

We can ask, whether there is a definition of the arrival time, which can be
valid both in classical and quantum mechanics. In our opinion, the words ``the
particle arrives from the left at the point $X$ at the time $t$'' mean that:
(i) at time $t$ the particle has been in the region $x<X$ and
(ii) at time $t+\Delta t$ ($\Delta t\rightarrow 0$) the particle has been found
in the region $x>X$.
Now we apply this definition, given by (i) and (ii), to the time of arrival in
the classical case.

Since quantum mechanics deals with probabilities, it is convenient to use
probabilistic description of the classical mechanics, as well. Therefore, we
will consider an ensemble of non interacting classical particles.  The
probability density in the phase space is $\rho(x,p;t)$.

Let us denote the region $x<X$ as $\Gamma_{1}$ and the region $x>X$ as
$\Gamma_{2}$. The probability that the particle arrives from the region
$\Gamma_{1}$ to the region $\Gamma_{2}$ at the time between $t$ and $t+\Delta t$
is proportional to the probability that the particle is in the region
$\Gamma_{1}$ at time $t$ and it is in the region $\Gamma_{2}$ at time $t+\Delta
t$. This probability is
\begin{equation}
\Pi_{+}(t)\Delta t=\frac{1}{N_{+}}\int_{\Omega}dpdx\rho(x,p;t),
\label{eq:p1}
\end{equation}
where $N_{+}$ is the constant of normalization and the region of the phase space
$\Omega$ has the following properties: (i) the coordinates of the points in
$\Omega$ are in the space region $\Gamma_{1}$ and (ii) if the phase trajectory
goes through the point of the region $\Omega$ at time $t$ then the particle at
time $t+\Delta t$ is in the space region $\Gamma_{2}$.  Since $\Delta t$ is
infinitesimal, the change of the coordinate during the time interval $\Delta t$
is equal to $\frac{p}{m}\Delta t$. Therefore, the particle arrives from the
region $\Gamma_{1}$ to the region $\Gamma_{2}$ only if the momentum of the
particle in the point $X$ is positive. The phase space region $\Omega$ consists
from the points with positive momentum $p$ and with coordinates between
$X-p/m\Delta t$ and $X$. Then from Eq.~(\ref{eq:p1}) we have the probability of
arrival time
\begin{equation}
\Pi_{+}(t)\Delta t=\frac{1}{N_{+}}\int_{0}^{\infty}dp
\int_{X-\frac{p}{m}\Delta t}^{X}dx\rho(x,p;t).
\label{eq:p2}
\end{equation}
Since $\Delta t$ is infinitesimal and the momentum of every particle is
finite, we can replace $x$ in Eq.~(\ref{eq:p2}) by $X$ and obtain the equality
\begin{equation}
\label{eq:toaprob}
\Pi_{+}(t,X)=\frac{1}{N_{+}}\int^{\infty}_{0}dp\frac{p}{m}\rho(X,p;t).
\end{equation}
The obtained arrival time distribution $\Pi_+(t,X)$ is well-known and has
appeared quite often in the literature (see, e.g., review \cite{review} and
references therein).

The probability current in classical mechanics is
\begin{equation}
\label{eq:flux}
J(x;t)=\int_{-\infty}^{+\infty}\frac{p}{m}\rho(x,p;t)dp .
\end{equation}
From Eqs.~(\ref{eq:toaprob}) and (\ref{eq:flux}) it is clear that the time of
arrival is related to the probability current. This relation, however, is not
straightforward. We can introduce the ``positive probability current''
\begin{equation}
\label{eq:positflux}
J_{+}(x;t)=\int_{0}^{\infty}\frac{p}{m}\rho(x,p;t)dp
\end{equation}
and rewrite Eq.~(\ref{eq:toaprob}) as
\begin{equation}
\label{eq:toaflux1}
\Pi_{+}(t,X)=\frac{1}{N_{+}}J_{+}(X;t) .
\label{eq:7a}
\end{equation}
The proposed \cite{Allcock,Bracken,Muga2} various quantum versions of $J_+$ even
in the case of the free particle can be negative (the so-called backflow
effect). Therefore, classical expression (\ref{eq:7a}) for the time of arrival
becomes problematic in quantum mechanics.

Similarly, for the arrival from the right we obtain the probability density
\begin{equation}
\label{eq:toaflux2}
\Pi_{-}(t,X)=\frac{1}{N_{-}}J_{-}(X;t) ,
\end{equation}
where the negative probability current is
\begin{equation}
J_{-}(x;t)=\int_{-\infty}^{0}\frac{|p|}{m}\rho(x,p;t)dp .
\end{equation}

We see that our definition, given at the beginning of this section, leads to the
proper result in the classical mechanics. The conditions (i) and (ii) does not
involve the concept of the trajectories. We can try to use this definition
also in quantum mechanics.

\section{\label{sec:weakmeas}Weak measurement}

The proposed definition of the arrival time probability distribution can be used
in the quantum mechanics only if the determination of the region, in which the
particle is, does not disturb the motion of the particle. This can be achieved
using the weak measurements of Aharonov, Albert and Vaidman
\cite{Aharonov2,Aharonov1,Duck,Aharonov3,Aharonov4,Aharonov5}.

We have the detector in the initial state $|\Phi\rangle$.  The detector
interacts with the particle only in the region $\Gamma_{1}$.  In order the weak
measurements can provide the meaningful information, the measurements must be
performed on an ensemble of identical systems. Each system with its own detector
is prepared in the same initial state. The readings of the detectors are
collected and averaged.

We take the operator describing the interaction between the particle and the
detector of the form \cite{Ruseckas1,Ruseckas2,Ruseckas3}
\begin{equation}
\label{eq:inter}
\hat{H}_{I}=\lambda\hat{q}\hat{P}_{1},
\label{eq:11a}
\end{equation}
where $\hat{P}_{1}$ is the projection operator, projecting into the region
$\Gamma_{1}$ and $\lambda$ characterizes the strength of the interaction with
the detector.  The interaction operator (\ref{eq:inter}) only slightly differs
from the one used by Aharonov, Albert and Vaidman \cite{Aharonov1}.  The
similar interaction operator has been considered by
von Neumann \cite{vNeum} and has been widely used in the strong
measurement models (e.g., \cite{Aharonov6,joos,caves,milb,gagen,Ruseckas4} and many others).

The measurement time is $\tau$. We assume that the interaction strength
$\lambda$ and the time $\tau$ are small. A very small parameter $\lambda$
ensures the undisturbance of the particle's evolution. The operator $\hat{q}$
acts in the Hilbert space of the detector. We require a continuous spectrum of
the operator $\hat{q}$. For simplicity, we can consider this operator as the
coordinate of the detector. The momentum, conjugate to $q$, is $p_q$.

Since interaction strength $\lambda$ and the duration of the measurement $\tau$
are small, the probability $\langle\hat{P}_{1}\rangle$ to find the particle
in the region $\Gamma_{1}$ does not change significantly during the measurement.
The action of the Hamiltonian (\ref{eq:inter}) results in the small change of
the mean detector's momentum
$\langle\hat{p}_{q}\rangle-\langle\hat{p}_{q}\rangle_0=-\lambda\tau
\langle\hat{P}_{1}\rangle$, where $\langle\hat{p}_{q}\rangle_0=
\langle\Phi(0)|\hat{p}_q|\Phi(0)\rangle$ is the mean
momentum of the detector at the beginning of the measurement and
$\langle\hat{p}_{q}\rangle=\langle\Phi(\tau)|\hat{p}_q|\Phi(\tau)\rangle$ is the
mean momentum of the detector after the measurement. Therefore, in analogy to
Ref.~\cite{Aharonov1}, we define the ``weak value'' of the probability to find
the particle in the region $\Gamma_{1}$,
\begin{equation}
W(1)\equiv\langle \hat{P}_{1}\rangle =\frac{\langle\hat{p}_{q}\rangle_{0}
-\langle\hat{p}_{q}\rangle}{\lambda\tau }.
\label{eq:defin}
\end{equation}

In order to obtain the arrival time probability using the definition from
Sec.~\ref{sec:classic}, we measure the momenta $p_{q}$ of each detector after
the interaction with the particle.  After the time $\Delta t$ we perform the
final, postselection measurement on the particles of our ensemble and measure if
the particle is found in the region $\Gamma_{2}$.  Then we collect the outcomes
$p_{q}$ only for the particles found in the region $\Gamma_{2}$.

The projection operator projecting into the region $\Gamma_2$ is $\hat{P}_2$. In
the Heisenberg representation this operator is
\begin{equation}
  \tilde{P}_2(t)=\hat{U}(t)^{\dag}\hat{P}_2\hat{U}(t),
\end{equation}
where $\hat{U}$ is the evolution operator of the free particle.  After the
measurement the state of the particle and the detector is
$\hat{U}_{M}(\tau)|\Phi\rangle |\Psi\rangle$, where $|\Psi\rangle$ is the
initial state of the particle and $\hat{U}_{M}$ is the evolution operator
of the particle interacting with the detector. The joint probability that the
detector has the momentum $p_{q}$ and the particle after the time $\Delta t$ is
found in the region $\Gamma_{2}$ is
\begin{equation}
W(p_{q},2)=\langle\Psi |\langle\Phi
|\hat{U}_{M}(\tau)^{\dag}|p_{q}\rangle\langle
 p_{q}|\tilde{P}_2(\Delta t)\hat{U}_{M}(\tau)|\Phi\rangle |\Psi\rangle ,
\end{equation}
where $|p_{q}\rangle$ is the eigenfunction of the momentum operator
$\hat{p}_{q}$, $\hat{P}_{2}$ is the projection operator, projecting into the
region $\Gamma_{2}$.  In quantum mechanics the probability that two
quantities simultaneously have definite values does not always exist. If the
joint probability does not exist then the concept of the conditional probability
is meaningless.  However, in our case operators $\hat{P}_{2}$ and
$|p_{q}\rangle\langle p_{q}|$ act in different spaces and commute, therefore,
the joint probability $W(p_{q},2)$ exists.

Let us define the conditional probability, i.e., the probability that the
momentum of the detector is $p_{q}$, provided that the particle after the time
$\Delta t$ is found in the region $\Gamma_{2}$.  This probability is given
according to the Bayes's theorem
\begin{equation}
W(p_{q}|2)=\frac{W(p_{q},2)}{W(2)},
\end{equation}
where $W(2)$ is the probability that the particle after the time $\Delta t$ is
found in the region $\Gamma_{2}$. The average momentum of the detector on
condition that the particle after the time $\Delta t$ is found in the region
$\Gamma_{2}$ equals to
\begin{equation}
\langle\hat{p}_{q}\rangle_{2}=\int p_{q}W(p_{q}|2)dp_{q}=\frac{1}{W(2)}\langle\Psi |
\langle\Phi |\hat{U}_{M}(\tau)^{\dag}\hat{p}_{q}\tilde{P}_{2}(\Delta t)
\hat{U}_{M}(\tau)|\Phi\rangle |\Psi\rangle ,
\label{eq:pq2}
\end{equation}
where $\tilde{P}_{2}(\Delta t)=\hat{U}(\Delta t)^{\dag}\hat{P}_{2}\hat{U}(\Delta
t)$ is the operator $\hat{P}_{2}$ in the Heisenberg picture. According to
Eq.~(\ref{eq:defin})
\begin{equation}
W(1|2)=\frac{\langle\hat{p}_{q}\rangle_{0}-\langle\hat{p}_{q}
\rangle_{2}}{\lambda\tau }=\frac{\langle\hat{p}_{q}\rangle_{0}W(2)
-\langle\hat{p}_{q}\rangle_{2}W(2)}{\lambda\tau W(2)} 
\label{eq:17}
\end{equation}
is the weak value of probability to find the particle in the region $\Gamma_{1}$
on condition that the particle after the time $\Delta t$ is in the region
$\Gamma_{2}$.  The probability that the particle is in the region $\Gamma_{1}$
and after the time $\Delta t$ is in the region $\Gamma_{2}$ equals to
\begin{equation}
W(1,2)=W(2)W(1|2).
\label{eq:18}
\end{equation}

When the measurement time $\tau$ is sufficiently small, the influence of the
Hamiltonian of the particle can be neglected and the evolution operator of the
particle and the detector can be expressed as
\[
\hat{U}_{M}(\tau)=\exp\left(-\frac{i}{\hbar}\lambda\tau
\hat{q}\hat{P}_{1}\right).
\]
We expand the operator $\hat{U}_{M}(\tau)$ into the series of the parameter
$\lambda$, assuming that $\lambda$ is small.  In the first-order approximation,
using Eqs.~(\ref{eq:17}) and (\ref{eq:18}), we obtain
\begin{equation}
W(1,2) \approx \frac{1}{2}\langle\tilde{P}_{2}(\Delta t)
\hat{P}_{1}+\hat{P}_{1}\tilde{P}_{2}(\Delta t)\rangle 
+\frac{i}{\hbar}\left(\langle\hat{p}_{q}\rangle\langle 
\hat{q}\rangle -\re\langle \hat{q}\hat{p}_{q}\rangle\right)
\langle [\hat{P}_{1},\tilde{P}_{2}(\Delta t)]\rangle .
\label{eq:w12} 
\end{equation}

The probability $W(1,2)$ is constructed using the conditions (i) and (ii) from
Sec.~\ref{sec:classic}: the weak measurement is performed to determine if the
particle is in the region $\Gamma_1$ and after the time $\Delta t$ the strong
measurement determines if the particle is in the region $\Gamma_2$. Therefore,
according to Sec.~\ref{sec:classic}, the quantity $W(1,2)$ after normalization
can be considered as the weak value of the arrival time probability
distribution.

Eq.~(\ref{eq:w12}) consists of two terms and we can introduce two quantities
\begin{equation}
\label{eq:20}
\Pi^{(1)}=\frac{1}{2\Delta t}\langle\hat{P}_{1}\tilde{P}_{2}(\Delta t)
+\tilde{P}_{2}(\Delta t)\hat{P}_{1}\rangle
\end{equation}
and
\begin{equation}
\Pi^{(2)}=\frac{1}{2i\Delta t}\langle [\hat{P}_{1},\tilde{P}_{2}(\Delta t)]
\rangle .
\end{equation}
Then
\begin{equation}
\label{eq:prob12}
W(1,2)=\Pi^{(1)}\Delta t-\frac{2\Delta t}{\hbar}\left(\langle\hat{p}_{q}
\rangle\langle \hat{q}\rangle -\re\langle \hat{q}\hat{p}_{q}\rangle\right)
\Pi^{(2)}.
\end{equation}

If the commutator $[\hat{P}_{1},\tilde{P}_{2}(\Delta t)]$ in
Eqs.~(\ref{eq:20})--(\ref{eq:prob12}) is not zero then, even in the limit of the
very weak measurement, the measured value depends on the particular detector.
This fact means that in such a case we cannot obtain the definite value for the
arrival time probability. Moreover, the coefficient
$(\langle\hat{p}_{q}\rangle\langle \hat{q}\rangle -\re\langle
\hat{q}\hat{p}_{q}\rangle)$ may be zero for the specific initial state of the
detector, e.g., for the Gaussian distribution of the coordinate $q$ and momentum
$p_{q}$.

The quantities $W(1,2)$, $\Pi^{(1)}$ and $\Pi^{(2)}$ are real. However, it is
convenient to consider the complex quantity
\begin{equation}
\label{eq:complex}
\Pi_{C}=\Pi^{(1)}+i\Pi^{(2)}=\frac{1}{\Delta t}\langle
\hat{P}_{1}\tilde{P}_{2}(\Delta t)\rangle .
\end{equation}
We name it ``complex arrival probability''. We can introduce the
corresponding operator
\begin{equation}
\hat{\Pi}_{+}=\frac{1}{\Delta t}\hat{P}_{1}\tilde{P}_{2}(\Delta t).
\end{equation}
By analogy, the operator
\begin{equation}
\hat{\Pi}_{-}=\frac{1}{\Delta t}\hat{P}_{2}\tilde{P}_{1}(\Delta t).
\end{equation}
corresponds to the arrival from the right.

The introduced operator $\hat{\Pi}_{+}$ has some properties of the classical
positive probability current. From the conditions $\hat{P}_{1}+\hat{P}_{2}=1$
and $\tilde{P}_{1}(\Delta t)+\tilde{P}_{2}(\Delta t)=1$ we have
\[
\hat{\Pi}_{+}-\hat{\Pi}_{-}=\frac{1}{\Delta t}(\tilde{P}_{2}(\Delta t)-\hat{P}_{2}).
\]
In the limit $\Delta t\rightarrow 0$ we obtain the probability current,
$\hat{J}=\lim_{\Delta t\rightarrow 0}(\hat{\Pi}_{+}-\hat{\Pi}_{-})$, as it is in
the classical mechanics. However, the quantity $\langle\hat{\Pi}_{+}\rangle$ is
complex and the real part can be negative, in contrary to the classical quantity
$J_{+}$. The reason for this is the non-commutativity of the operators
$\hat{P}_{1}$ and $\tilde{P}_{2}(\Delta t)$.  When the imaginary part is small,
the quantity $\langle\hat{\Pi}_{+}\rangle$ after normalization can be considered
as the approximate probability distribution of the arrival time .

\section{\label{sec:free}Arrival time probability}

The operator $\hat{\Pi}_{+}$ is obtained without specification of the
Hamiltonian of the particle and is suitable to the free particles and to the
particles subjected to an external potential as well. In this section we
consider the arrival time of the free particle.

The calculation of the ``weak arrival time distribution'' $W(1,2)$ involves the
average $\langle\hat{\Pi}_{+}\rangle$. Therefore, it is useful to have the
matrix elements of the operator $\hat{\Pi}_{+}$. It should be noted, that matrix
elements of the operator $\hat{\Pi}_{+}$ as well as the operator itself are only
auxiliary and does not have independent meaning.

In the basis of momentum eigenstates $|p\rangle$, normalized according the
condition $\langle p_1|p_2 \rangle=2\pi\hbar\delta(p_1 - p_2)$, the matrix
elements of the operator $\hat{\Pi}_{+}$ are
\begin{eqnarray}
\langle p_{1}|\hat{\Pi}_{+}|p_{2}\rangle  & = & \frac{1}{\Delta t}\langle
p_{1}|\hat{P}_{1}
\hat{U}(\Delta t)^{\dag}\hat{P}_{2}\hat{U}(\Delta t)|p_{2}\rangle\nonumber\\
 & = & \frac{1}{\Delta t}\int^{X}_{-\infty}dx_{1}\int^{\infty}_{X}
dx_{2}e^{-\frac{i}{\hbar}p_{1}x_{1}}\langle x_{1}|\hat{U}(\Delta t)^{\dag}|x_{2}
\rangle e^{\frac{i}{\hbar}p_{2}x_{2}-\frac{i}{\hbar}
\frac{p^{2}_{2}}{2m}\Delta t} .
\end{eqnarray}
After performing the integration we obtain
\begin{eqnarray}
\langle p_{1}|\hat{\Pi}_{+}|p_{2}\rangle  & = & \frac{i\hbar}{2\Delta t
(p_{2}-p_{1})}
\exp\left(\frac{i}{\hbar}(p_{2}-p_{1})X\right)\nonumber\\
 &  & \times\left(e^{\frac{i}{\hbar}\frac{\Delta t}{2m}(p_{1}^{2}-p^{2}_{2})}
\erfc\left(-p_{1}\sqrt{\frac{i\Delta t}{2\hbar m}}\right)-
\erfc\left(-p_{2}\sqrt{\frac{i\Delta t}{2\hbar m}}\right)\right) ,
\label{eq:p1p2} 
\end{eqnarray}
where $\sqrt{i}=\exp(i\pi/4)$. When
\[
\frac{1}{\hbar}\frac{\Delta t}{2m}(p_{1}^{2}-p^{2}_{2})\ll 1,
\quad p_{1}\sqrt{\frac{\Delta t}{2\hbar m}}>1,
\quad p_{2}\sqrt{\frac{\Delta t}{2\hbar m}}>1,
\]
the matrix elements of the operator $\hat{\Pi}_{+}$ are
\begin{equation}
\langle p_{1}|\hat{\Pi}_{+}|p_{2}\rangle\approx \frac{p_{1}+p_{2}}{2m}
\exp\left(\frac{i}{\hbar}(p_{2}-p_{1})X\right) .
\end{equation}
This equation coincides with the expression for the matrix elements of the
probability current operator.

From Eq.~(\ref{eq:p1p2}) we obtain the diagonal matrix elements of the operator
$\hat{\Pi}_{+}$
\begin{equation}
\label{eq:diagonal}
\langle p|\hat{\Pi}_{+}|p\rangle =\frac{p}{2m}
\erfc\left(-p\sqrt{\frac{i\Delta t}{2\hbar m}}\right)+
\frac{\hbar}{\sqrt{i2\pi\hbar m\Delta t}}e^{-\frac{i}{\hbar}
\frac{p^{2}}{2m}\Delta t} .
\end{equation}

\begin{figure}
\begin{center}
\includegraphics{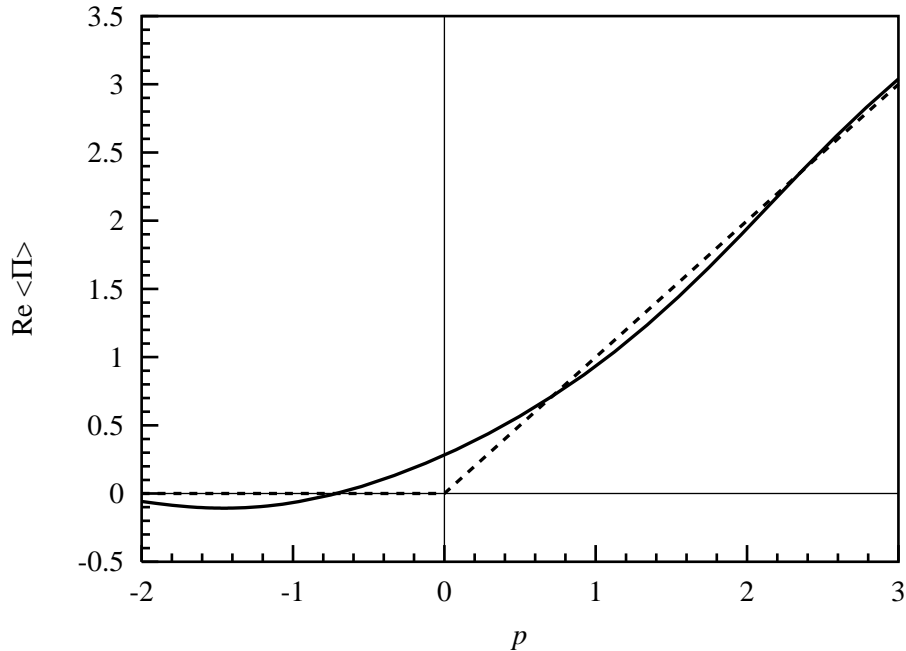}
\end{center}
\caption{The real part of the quantity
  \protect$\langle p|\hat{\Pi}_{+}|p\rangle\protect$, according to
  Eq.~(\ref{eq:diagonal}). The corresponding classical positive probability
  current is shown with the dashed line. The used parameters are \protect$\hbar
  =1\protect$, \protect$m=1\protect$, and \protect$\Delta t=1\protect$. In this
  system of units, the momentum \protect$p\protect$ is dimensionless.}
\label{fig:1}
\end{figure}

\begin{figure}
\begin{center}
\includegraphics{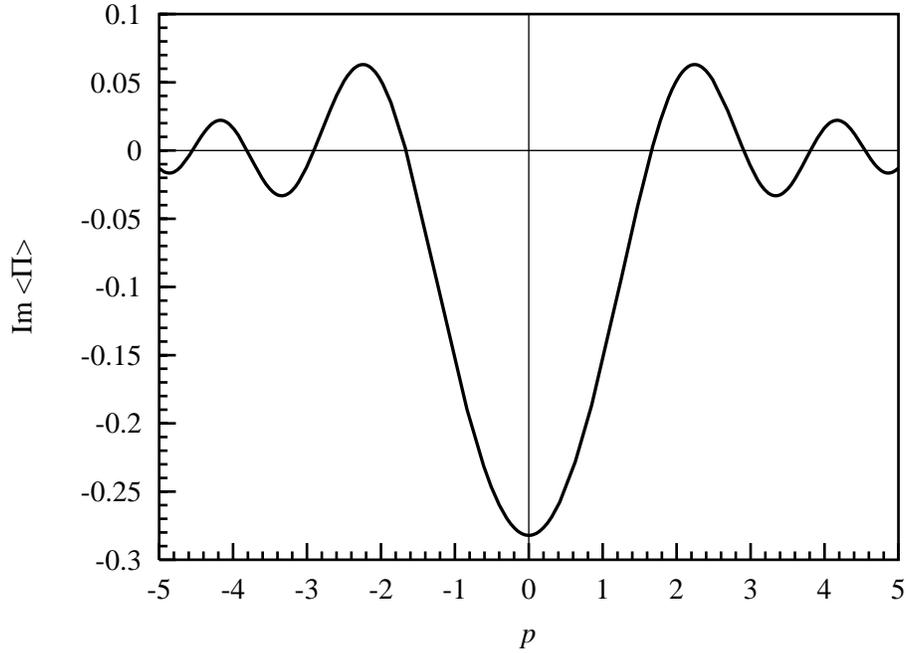}
\end{center}
\caption{The imaginary part of the quantity
  \protect$\langle p|\hat{\Pi}_{+}|p\rangle\protect$.  Used parameters are the
  same as in Fig. \ref{fig:1}}
\label{fig:2}
\end{figure}
The real part of the quantity $\langle p|\hat{\Pi}_{+}|p\rangle$ is shown in
Fig.~\ref{fig:1} and the imaginary part in Fig.~\ref{fig:2}.

Using the asymptotic expressions of the function $\erfc$ we obtain from
Eq.~(\ref{eq:diagonal}) that
\[
\lim_{p\rightarrow +\infty}\langle p|\hat{\Pi}_{+}|p\rangle\rightarrow
\frac{p}{m}
\]
and $\langle p|\hat{\Pi}_{+}|p\rangle\rightarrow 0$, when $p\rightarrow
-\infty$, i.e., the imaginary part tends to zero and the real part approaches
the corresponding classical value as the modulus of the momentum $|p|$
increases. Such a behaviour is evident from Figs.~\ref{fig:1} and \ref{fig:2},
too.

The asymptotic expressions of the function $\erfc$ are valid when the argument
of the $\erfc$ is large, i.e., $|p|\sqrt{\frac{\Delta t}{2\hbar m}}>1$ or
\begin{equation}
\label{eq:cond}
\Delta t>\frac{\hbar}{E_{k}} .
\end{equation}
Here $E_{k}$ is the kinetic energy of the particle.
\begin{figure}
\begin{center}
\includegraphics{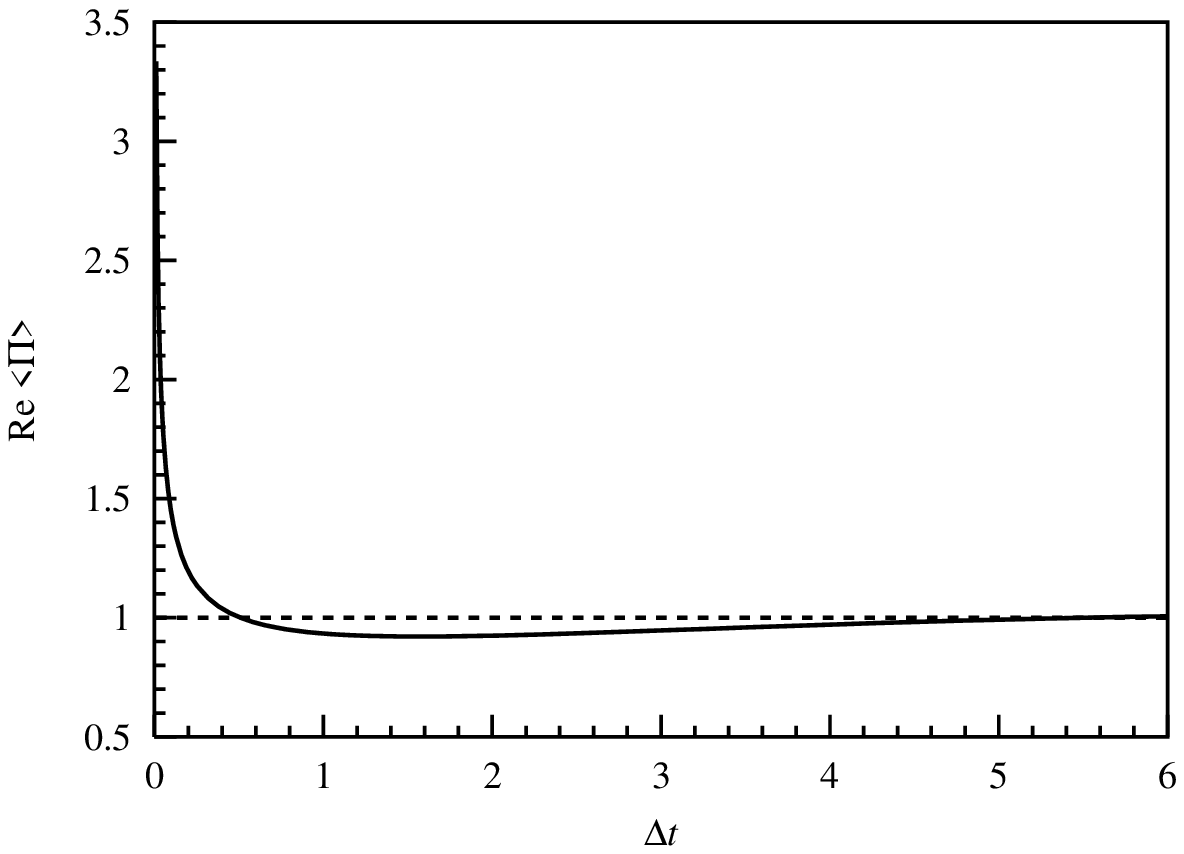}
\end{center}
\caption{The dependence of the quantity
  \protect$\re\langle p|\hat{\Pi}_{+}|p\rangle\protect$ according to
  Eq.~(\ref{eq:diagonal}) from the resolution time \protect$\Delta t\protect$.
  The corresponding classical positive probability current is shown with the
  dashed line. The used parameters are \protect$\hbar =1\protect$,
  \protect$m=1\protect$, and \protect$p=1\protect$. In these units,
  the time \protect$\Delta t\protect$ is dimensionless.}
\label{fig:3}
\end{figure}
The dependence of the quantity $\re\langle p|\hat{\Pi}_{+}|p\rangle$ from
$\Delta t$ is shown in Fig.~\ref{fig:3}. For small $\Delta t$ the quantity
$\langle p|\hat{\Pi}_{+}|p\rangle$ is proportional to $1/\sqrt{\Delta t}$.
Therefore, unlike the classical mechanics, in quantum mechanics $\Delta t$
cannot be zero. Eq.~(\ref{eq:cond}) imposes the lower bound to the resolution
time $\Delta t$. It follows that our model does not permit determination of the
arrival time with the resolution greater than $\hbar /E_{k}$. The relation,
similar to Eq.~(\ref{eq:cond}), based on measurement models has been obtained by
Aharonov \emph{et al.\/} \cite{Aharonov6}. The time-energy uncertainty
realtions, associated with time of arrival distribution, are also discussed
in Refs.~\cite{Baute4,Brunetti2}

\section{\label{sec:concl}Conclusion}

The definition of density of one sided arrivals is proposed. This definition is
extended into quantum mechanics, using the concept of weak measurements by
Aharonov \emph{et al.} The proposed procedure is suitable to the free
particles and to the particles subjected to an external potential, as well. It
gives not only the mathematical expression for the arrival time probability
distribution but also the way to measure the obtained quantity. However, this
procedure gives no unique expression for the arrival time probability
distribution.

In analogy with complex tunneling time the complex arrival time ``probability
distribution'' is introduced (Eq.~(\ref{eq:complex})).
It is shown that the proposed approach imposes an inherent
limitation, Eq.~(\ref{eq:cond}), to the resolution time of the arrival time
determination.


\end{document}